# REVERSIBLE LOGIC SYNTHESIS OF FAULT TOLERANT CARRY SKIP BCD ADDER

MD. SAIFUL ISLAM AND ZERINA BEGUM

*Institute of Information Technology, University of Dhaka, Dhaka-1000, Bangladesh*

**ABSTRACT**

Reversible logic is emerging as an important research area having its application in diverse fields such as low power CMOS design, digital signal processing, cryptography, quantum computing and optical information processing. This paper presents a new 4*4 parity preserving reversible logic gate, IG. The proposed parity preserving reversible gate can be used to synthesize any arbitrary Boolean function. It allows any fault that affects no more than a single signal readily detectable at the circuit's primary outputs. It is shown that a fault tolerant reversible full adder circuit can be realized using only two IGs. The proposed fault tolerant full adder (FTFA) is used to design other arithmetic logic circuits for which it is used as the fundamental building block. It has also been demonstrated that the proposed design offers less hardware complexity and is efficient in terms of gate count, garbage outputs and constant inputs than the existing counterparts.

## INTRODUCTION

Power dissipation is an important factor in VLSI design. Combinational logic circuits dissipate heat in an order of $kT \ln 2$ joules for every bit of information that is lost, where $k$ is the Boltzman constant and T is the operating temperature [1]. Information is lost when the input vector cannot be uniquely recovered from its output vectors. Reversible logic circuits naturally take care of heating since in a reversible logic every input vector can be uniquely recovered from its output vectors and therefore no information is lost. According to [2] zero energy dissipation would be possible only if the network consists of reversible gates. Thus reversibility will become an essential property in future circuit design.

Synthesis of reversible logic circuits differs from the combinational one in many ways [3]. Firstly, in reversible circuit there should be no fan-out, that is, each output will be used only once. Secondly, for each input pattern there should be a unique output pattern. Finally, the resulting circuit must be acyclic. Any reversible gate performs the permutation of its input patterns only and realizes the functions that are reversible. If a reversible gate has $k$ inputs, and therefore $k$ outputs, then we call it a $k*k$ reversible gate. Any reversible circuit design includes only the gates that are reversible. In a reversible circuit, the outputs that are not used as primary outputs are called garbages and the input lines that are set to constants are termed as constant inputs. An efficient design should keep both the number of garbage outputs and constant inputs to minimum.

Fault tolerance is the property that enables a system to continue operating properly in the event of the failure of some its components. If the system itself made of fault tolerant components, then the detection and correction of faults become easier and simple. In communication and many other systems, fault tolerance is achieved by parity. Therefore, parity preserving reversible circuits will be the future design trends towards the development of fault tolerant reversible systems in nanotechnology and a gating network will be parity preserving if its individual gate is parity preserving [4]. Thus, we need parity preserving reversible logic gates to construct parity preserving reversible circuits. This paper presents a new 4*4 parity preserving logic gate, IG. It is parity preserving, that is, the parity of the inputs matches the parity of the outputs. IG is universal in the sense that it can be used to synthesize any arbitrary Boolean function. It is shown that a fault tolerant reversible full adder circuit can be realized using only two IGs. The presented design does not produce any unnecessary garbage outputs. Minimizing number garbage outputs are the major concern in reversible logic design [3]. The presented FTFA block can be used to realize other fault tolerant arithmetic logic circuits in nanotechnology such as ripple carry adder, carry look-ahead adder, carry-skip logic, and multiplier/divisors.

## REVERSIBLE LOGIC GATES

1. *Basic Reversible Gates*: There exist many reversible gates in the literature. Among them 2*2 Feynman gate (FG) [6], depicted in Fig. 1a, 3*3 Peres gate (PG) [7], depicted in Fig. 1b, 3*3 Toffoli gate (TG) [8], depicted in Fig. 1c and 3*3 Fredkin gate (FRG) [9], depicted in Fig. 1d have been studied extensively. Because of their simplicity and quantum realization cost there are design approaches and tools that incorporate them separately or in combination with each other [3] [5].

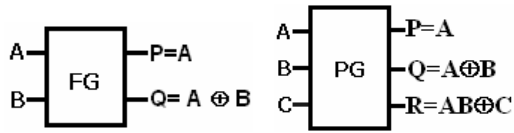

Fig. 1 a. 2*2 Feynman gate.     Fig. 1 b. 3*3 Peres gate.

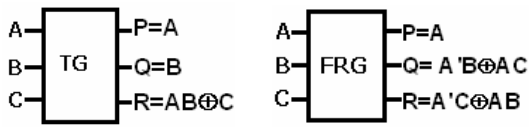

Fig. 1 c. 3*3 Toffoli gate.     Fig. 1 d. 3*3 Fredkin gate.

2. *Parity Preserving Reversible Gates:* A reversible gate is called parity preserving reversible gate if its input parity matches the parity of its output. More formally, any k*k reversible logic gate where the EX-OR of the inputs matches the EX-OR of the outputs will be parity preserving. A few parity preserving logic gates have been proposed in the literature. Among them 3*3 Feynman Double gate (F2G) [4] depicted in Fig. 2a and 3*3 Fredkin gate (FRG) [9] depicted in Figure 2b are one-through gates, which means one of the inputs is also output. Recently a new 3*3 parity preserving reversible gate, namely New Fault Tolerant gate (NFT) depicted in Fig. 2c [10] and a 4*4 parity preserving HC gate (PPHCG) [11] depicted in Fig. 2d have been proposed.

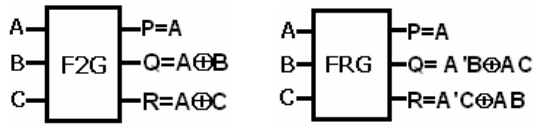

Fig. 2 a. 3*3 Feynman Double gate.    Fig. 2 b. 3*3 Fredkin gate.

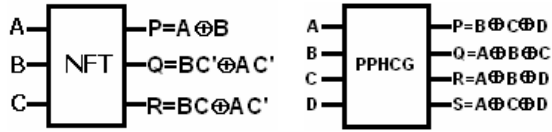

Fig. 2 c. 3*3 NFT gate.    Fig. 2 d. 4*4 PPHCG gate.

3. *A New 4*4 Parity Preserving Reversible Gate:* This paper presents a new 4*4 parity preserving reversible gate, IG, depicted in Fig. 3. The gate is one-through, which means one of the input variables is also output. The corresponding truth table of the gate is shown in Table 1. It can be verified from the truth table that the input pattern corresponding to particular output pattern can be uniquely determined. The proposed reversible IG is parity preserving. This is readily verified by comparing the input parity A⊕B⊕C⊕D to the output parity P⊕Q⊕R⊕S. The newly proposed IG gate is universal in the sense that it can be used for implementing arbitrary Boolean functions as shown in Fig. 4.

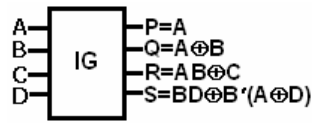

Fig. 3. A new 4*4 parity preserving reversible gate IG.

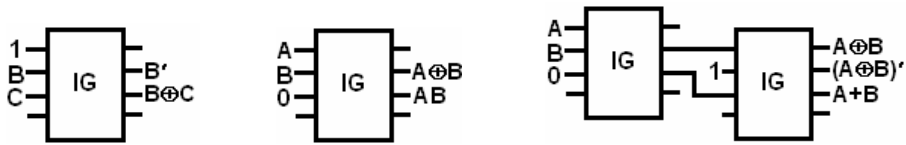

a. IG as Inverter and EX-OR. b. IG as AND and EX-OR. c. IG as EXOR, EX-NOR and OR

Fig. 4. Proposed IG gate can implement all Boolean functions.

TABLE 1.

TRURTH TABLE OF THE PROPOSED PARITY PRESERVING IG GATE

| A | B | C | D | P | Q | R | S |
|---|---|---|---|---|---|---|---|
| 0 | 0 | 0 | 0 | 0 | 0 | 0 | 0 |
| 0 | 0 | 0 | 1 | 0 | 0 | 0 | 1 |
| 0 | 0 | 1 | 0 | 0 | 0 | 1 | 0 |
| 0 | 0 | 1 | 1 | 0 | 0 | 1 | 1 |
| 0 | 1 | 0 | 0 | 0 | 1 | 0 | 0 |
| 0 | 1 | 0 | 1 | 0 | 1 | 0 | 1 |
| 0 | 1 | 1 | 0 | 0 | 1 | 1 | 0 |
| 0 | 1 | 1 | 1 | 0 | 1 | 1 | 1 |
| 1 | 0 | 0 | 0 | 1 | 1 | 0 | 1 |
| 1 | 0 | 0 | 1 | 1 | 1 | 0 | 0 |
| 1 | 0 | 1 | 0 | 1 | 1 | 1 | 1 |
| 1 | 0 | 1 | 1 | 1 | 1 | 1 | 0 |
| 1 | 1 | 0 | 0 | 1 | 0 | 1 | 0 |
| 1 | 1 | 0 | 1 | 1 | 0 | 1 | 1 |
| 1 | 1 | 1 | 0 | 1 | 0 | 0 | 0 |
| 1 | 1 | 1 | 1 | 1 | 0 | 0 | 1 |

**FAULT TOLERANT FULL ADDER CIRCUIT**

Reversible logic implementation of full adder circuit has been studied by several authors in the literature [3-4], [12-13]. It has been demonstrated in [3] that a reversible full adder circuit can be realized with at least two garbage outputs and one constant input. This requirement is not the same for fault tolerant reversible full adder circuit. Because in a fault tolerant full adder circuit the input parity must matches the parity of the outputs. This section first establishes the minimum number of garbage outputs and constant inputs required to design a fault tolerant reversible full adder circuit. Then proposes a new realization of fault tolerant reversible full adder circuit using the newly proposed IG gates.

**Theorem 2:** Any realization of a fault tolerant reversible full adder circuit needs at least three garbage outputs and two constant inputs.

*Proof*: The full adder circuit output equations $S=A\oplus B\oplus C_{in}$ and $C_{out}= (A\oplus B)Cin\oplus AB$ produce the same output S=1 and $C_{out}$=0, for the three distinct input combinations A=0, B=0, $C_{in}$=1; A=0, B=1, $C_{in}$=0; and A=1, B=0, $C_{in}$=0. The parity of the input vector matches the parity of the corresponding output vector. To separate all repeated values of outputs S and $C_{out}$ as well as keeping their parity unchanged, we need at least three garbage outputs. Thus the total number of outputs is 2+3=5. Now since in a reversible circuit the number of inputs must be equal to the number of outputs and there are three inputs in a full adder circuit A, B and $C_{in}$, the other two inputs need to be constant inputs. ∎

TABLE 2.

INPUT COMBINATIONS THAT PRODUCE THE SAME OUTPUT COMBINATIONS IN FULL ADDER CIRCUIT (SHOWN SHADED)

| Input | | | | | Output | | | | |
|---|---|---|---|---|---|---|---|---|---|
| A | B | Cin | C1 | C2 | S | Cout | G1 | G2 | G3 |
| 0 | 0 | 1 | 0 | 0 | 1 | 0 | 0 | 0 | 0 |
| 0 | 1 | 0 | 0 | 0 | 1 | 0 | 0 | 1 | 1 |
| 1 | 0 | 0 | 0 | 0 | 1 | 0 | 1 | 0 | 1 |

There are two fault tolerant reversible full adder circuits in the literature [12], [13]. The fault tolerant full adder circuit in [12] requires six parity preserving reversible gates (two FRGs and four F2Gs) and the fault tolerant full adder circuit in [13] uses four FRGs. This paper presents a new design of fault tolerant reversible full adder circuit that uses only two IGs, depicted in Fig. 5. It requires only two clock cycles.

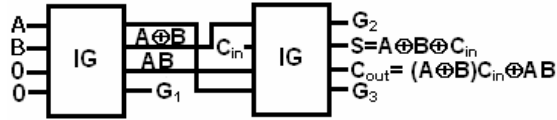

Fig. 5. Proposed fault tolerant reversible full adder circuit.

*Evaluation of the Proposed Full Adder Circuit*: Evaluation of the proposed FTFA can be comprehended easily with the help of the comparative results given in Table 3. One of the main factors of a circuit is its hardware complexity. It can be proved that the proposed circuit is better than the existing approaches in terms of hardware complexity. Let

$\alpha$ = A two input EXOR gate calculation

$\beta$ = A two input AND gate calculation

$\delta$ = A NOT gate calculation

T = Total logical calculation

For [13] the Total logical calculation is: $T=8\alpha+16\beta+4\delta$, for [12] the Total logical calculation is: $12\alpha+8\beta+2\delta$, and for our proposed fault tolerant reversible full adder circuit, the Total logical calculation is: $8\alpha+6\beta+2\delta$.

TABLE 3.

COMPERATIVE RESULTS OF DIFFERENT FAULT TOLERANT FULL ADDER CIRCUITS

| | Gate Count | Clock cycles | Garbage outputs | Constant Inputs | Total logical calculation |
|---|---|---|---|---|---|
| Proposed Circuit | 2 IGs=2 | 2 | 3 | 2 | $8\alpha+6\beta+2\delta$ |
| Existing circuit [13] | 4 FRGs =4 | 4 | 3 | 2 | $8\alpha+16\beta+4\delta$ |
| Existing circuit [12] | 2FRGs+4 F2Gs=6 | 6 | 6 | 5 | $12\alpha+8\beta+2\delta$ |

One of the other major constraints in designing a reversible logic circuit is to lessen number of garbage outputs. The proposed FTFA produces only three garbage outputs which are equal to the design in [13] and this is minimum as proved earlier in this paper, but the design in [12] produces six garbage outputs.

Number of constant inputs is one of the other main factors in designing a reversible logic circuit. The proposed FTFA requires only two constant inputs that are equal to the design in [13] and this is the minimum theoretically, but the design in [12] requires 5 constant inputs. So, it can be stated that the proposed design approach is better than all the existing designs in terms of number of constant inputs.

**FAULT TOLERANT RIPPLE CARRY ADDER**

The full adder is the basic building block in a ripple carry adder. The reversible ripple carry adder using the FTFAs is shown in Fig. 6, which is obtained by cascading the full adders in series. The output expressions for a ripple carry adder are:

$$S_i = A \wedge B \wedge C_i \qquad (1)$$
$$C_{i+1} = (A \wedge B) \cdot C_i \wedge AB \ (i=0, 1, 2\ldots) \qquad (2)$$

*Evaluation of the Proposed Ripple Carry Adder:* It can be inferred from Fig. 5 and Fig. 6 that for N bit addition; the proposed ripple carry adder architecture uses only 2N reversible IG gates and produces only 3N garbage outputs. Table 4 shows the results that compare the proposed ripple carry adder with those designed using full adders of [12-13]. It is observed that the proposed circuit is better than existing ones both in terms of gate count and garbage outputs.

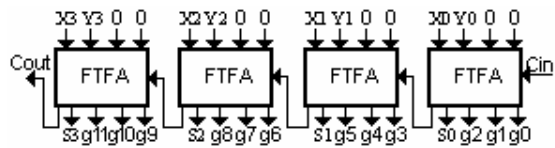

Fig. 6. Reversible logic implementation of fault tolerant ripple carry adder using the proposed FTFA.

TABLE 4.
COMPERATIVE RESULTS OF DIFFERENT REVERSIBLE CARRY ADDER

|  | Gate count | Garbage outputs produced | Constant inputs required | Unit delay |
|---|---|---|---|---|
| Proposed circuit | 2N | 3N | 2N | 2N |
| Existing circuit [13] | 4N | 3N | 2N | 4N |
| Existing circuit [12] | 6N | 6N | 5N | 6N |

# Fault Tolerant Carry Skip BCD Adder

The block diagram of the proposed fault tolerant carry-skip adder is shown in Fig. 7. It includes FTFA blocks presented earlier in this paper, Fredkin gates and parity preserving HC gates. The three Fredkin gates in the middle of the Fig. 7 are used to perform the AND4 operation. This generates the block propagate signal 'P'. The single FRG in the right side of Fig. 7 performs the AND-OR function to create the carry skip logic.

In the proposed carry skip adder, the FRG propagates the block's carry input '$C_{in}$' to the next block if the block propagate signal 'P' is one; otherwise, the most significant full adder carry '$C_4$' is propagated to the next block. The traditional carry skip AND-OR logic [14] [15] and the reversible carry skip logic in Fig. 7 do not have equivalent truth tables, but it must be noted that the Fredkin carry skip logic more faithfully adheres to the spirit of carry skip addition, by propagating the correct value of '$C_{in}$' to '$C_{out}$' [13][14]. Furthermore, it passes '$C_{in}$' to '$C_{out}$' whenever 'P'=1, regardless of the value of '$C_4$', thus time saving is significant.

It has been observed in [15] that the 3 inputs OR gate, used for generating the '$C_{out}$' can be replaced by 3 inputs XOR gate. To generate 3 inputs X-OR output, parity preserving HC gate has been used in the proposed carry skip logic rather than TS-3 gate [15] since TS-3 is not parity preserving. The proposed design includes 8 FTFA blocks, 6 FRGs and one PPHCG gate. The number of garbage outputs produced is 35. It should be noted that the proposed design produces more garbage outputs than the design presented in [15] because of the parity preservation.

*Evaluation of the Proposed Carry Skip BCD Adder:* The proposed design includes a total of 15 reversible units: 8 FTFA blocks, 6 FRGs and one PPHCG gate. The number of garbage outputs produced is 36. It should be noted that the proposed design produces more garbage outputs than the design presented in [15] because of the parity preservation. It has been proved earlier that a FTFA can be realized with no less than three garbage outputs and two constant inputs. Therefore, we can say that the proposed design is optimized in terms of garbage outputs and constant inputs.

# Conclusion

This paper presents a new 4*4 parity preserving reversible gate called IG gate and demonstrates its universality by realizing all possible Boolean functions. A novel fault tolerant reversible full adder circuit using the proposed IG gates has also been presented and optimized in terms of gate count, number of garbage outputs and constant inputs. Reversible logic implementation of optimized fault tolerant carry skip BCD adder has also been presented. The presented adder architectures using the proposed reversible gate offer less hardware complexity and optimized in terms of area and power consumption.

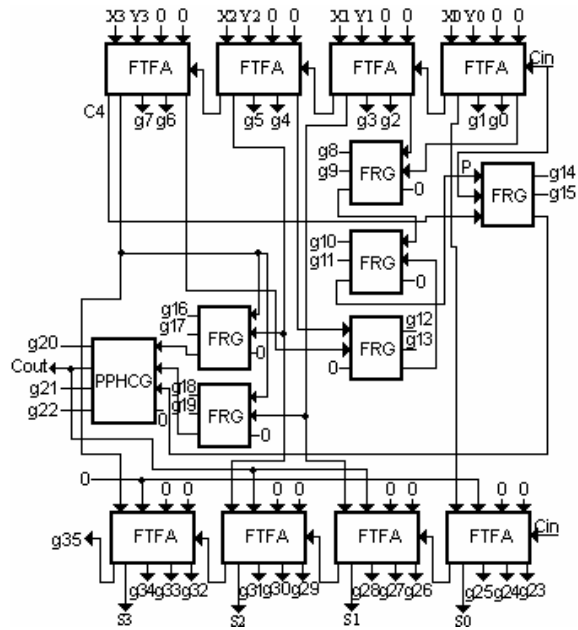

Fig 7. Reversible logic implementation of fault tolerant carry skip BCD adder.

TABLE 5.

COMPERATIVE ANALYSIS OF THE FAULT TOLERANT CARRY SKIP ADDER

|  | Reversible Units Used | Garbage outputs produced |
|---|---|---|
| Proposed circuit | 15 | 36 |
| Existing circuit [15] | 15 | 27 |